\documentstyle[pra,aps,epsf]{revtex}
\begin{document}
\draft
%\twocolumn
\title{Cooperative effects in the light and dark periods of two 
dipole-interacting atoms }
\author{ Almut Beige\thanks{Present address: 
 Blackett Laboratory, Imperial College London, 
London SW7 2BZ, England;
 email: a.beige@ic.ac.uk} 
} 
\address{Institut f\"ur Physik, Universit\"at Potsdam, D-14469 Potsdam,
  Germany}
\author{Gerhard C. Hegerfeldt\thanks{email: 
    hegerf@theorie.physik.uni-goettingen.de}}
\address{Institut f\"ur Theoretische Physik, Universit\"at G\"ottingen,
  Bunsenstrasse 9, D-37073 G\"ottingen, Germany}
\maketitle

\begin{abstract}  If an atom is able
  to exhibit macroscopic dark periods, or electron shelving, then a
  driven system of two atoms has three 
  types of fluorescence periods (dark, single and double
  intensity). We propose to use the average durations of these
  fluorescence types as a simple and easily accessible  
indicator of cooperative effects. As an
  example we study two dipole-interacting $V$ systems by simulation
  techniques. We show that  the durations of the two types of
  light periods exhibit  marked separation-dependent
  oscillations and that they vary in phase with the real part of the
  dipole-dipole coupling constant.
\end{abstract}
\pacs{PACS numbers: 42.50.Ar, 42.50.Fx}

\section{ Introduction}
Cooperative effects in the radiative behavior of atoms stored in a
trap may arise from their mutual dipole-dipole interaction if the
atoms are close enough to each other. This is particularly interesting for
two or three atoms and has attracted considerable interest in the
literature \cite{aga}-\cite{Ficek98}. Recently, the present
authors investigated in detail the transition from antibunching to
bunching for two two-level systems with decreasing atomic distance 
\cite{BeHe4}.

For a single multi-level system with a metastable state one has
the striking phenomenon of macroscopic dark periods, or electron
shelving, in which the electron is essentially shelved for
seconds or even minutes in a metastable state without photon emissions
\cite{SauterL}-\cite{HePle1}. For two such systems their
fluorescence behavior would, without cooperative effects, be just the
sum of the separate photon emissions, with dark periods of both atoms,
light periods of a single  and of two atoms. In Ref. \cite{Sauter} the
fluorescence intensity of two and three such ions in a Paul trap was
measured and a large fraction of almost simultaneous jumps by two and even
all three ions was recorded. This fraction was orders of magnitudes larger
than that expected for independent ions. A quantitative explanation of such a
large cooperative effect has been found to be difficult
\cite{hendriks,Java,Aga88,Lawande89,Chun}, and we will briefly
discuss this question again in the last section. Other experiments at
larger distances and with different ions showed no cooperative
effects  \cite{Itano88,Thom}.
A numerical approach to the study of double jumps faces the problem
that for good statistics one needs very long simulation times.

As a simpler indicator of cooperative effects for systems with light
and dark periods we propose here to use the mean durations, $T_0$,
$T_1$, and $T_2$, of the three
different types of fluorescence periods, i.e. dark,  single-intensity, and
double-intensity periods, respectively. The running time can be much
shorter than required for double jumps, making these quantities
easily accessible, both experimentally and in  simulations.

In this paper we therefore present a study of cooperative effects on
the mean durations of the three types of fluorescence periods of two
three-level $V$ systems with a metastable state, as a function of
their distance $r$. The level scheme of a single $V$ system is
depicted in Fig. 1 \cite{rem1}. Our simulations show that  the mean
durations of the single- and double-intensity periods, $T_1$ and $T_2$, 
depend sensitively on the dipole-dipole interaction and thus on the atomic
distance $r$. They exhibit marked oscillations which decrease in
amplitude when $r$ increases. These oscillations seem to continue up
to a distance of well over five wavelengths of the strong transition
where we have stopped our simulation. The real part of the
dipole-dipole coupling constant of the two $V$ systems also exhibits
damped oscillations. As a remarkable fact we find that these
oscillations are in phase with those of $T_1$ and $T_2$. This
correspondence is easy to understand intuitively since the real part
of the dipole-dipole coupling constant enters the decay rates of the
excited states of the combined two-atom system.

In Section II we explain the methods employed which are based on the
quantum jump approach \cite{HeWi}-\cite{PleKni} (equivalent to the
Monte-Carlo wavefunction approach \cite{MC} and to quantum
trajectories \cite{QT}). This approach is here adapted to two
dipole-interacting $V$ systems.

In Section III we define in more detail the three types of fluorescence
periods of zero, single and double intensity. This involves an
averaging procedure, both experimentally and theoretically. We then
present the results of our simulations.

In Section IV the numerical results are interpreted, and it is shown
that one can associate three distinct subspaces of states of the
coupled system with the three types of fluorescence periods. During
each period the coupled system remains in the corresponding
subspace. In Section V we discuss and summarize our results.
\section{The Quantum Jump Approach}
In the quantum jump approach \cite{HeWi,Wi,He,HeSo,PleKni}, the time
development 
of an atomic system is described by a conditional non-hermitian
Hamiltonian, $H_{{\rm cond}}$, which gives the time development between
photon emissions, and by a reset operation which gives the state or
density matrix right after an emission. For a general $N$-level system
these have been derived in Refs. \cite{He,HeSo}. The derivation
 is adapted here for a system consisting of two atoms. Slight
modifications arise here  since  the field operator 
appears with different position arguments.

We consider two atoms, each  a $V$ configuration as shown in
Fig. 1, with the levels $|1 \rangle_i$, $|2 \rangle_i$ and $|3 \rangle_i$ 
$(i=1,2)$ and fixed at positions ${\bf r}_i$. We define operators 
$S^\pm_{ij}$ $(i=1,2; ~j=2,3)$ in the two-atom Hilbert space by $S^+_{ij} = 
|j\rangle_{ii} \langle 1|$ and $S_{ij}^- = |1 \rangle_{ii} \langle j|$. 
For simplicity we consider the case where the  dipole moments of two
atoms are the same, i.e. $_1\langle 1|{\bf X}_1 |j\rangle_1 =~
_2\langle 1|{\bf X}_2 |j\rangle_2 \equiv {\bf D}_{1j}$. If ${\bf
  D}_{1j}$ is real then the  angle it forms
 with the line connecting the atoms is denoted by 
$\vartheta_j$. In general $\vartheta_j$ is defined through 
$\cos^2 \vartheta_j = \left| \left( {\bf D}_{1 j} , {\bf
        r} \right) \right|^2/r^2 D_{1j}^2$, where ${\bf r} = {\bf r}_2
  - {\bf r}_1$. We assume the 
laser radiation normal to this line so that the lasers are in 
phase for both atoms. The two lasers are  denoted by $L_2$ and $L_3$. 
We take zero detuning and 
${\bf E}_{Lj} ({\bf r},t) = {\rm Re}\left[ {\bf E}_{0j}~ \exp\{- i 
(\omega_{j1} t -{\bf k}\cdot {\bf r})\} \right]$ where 
$\hbar \omega_{j1}$ is the energy 
difference between level 1 and level $j$. Making the usual rotating-wave 
approximation and going over to the interaction picture the interaction 
Hamiltonian becomes
\begin{eqnarray} \label{21}
H_{\rm I} = \sum_{i=1}^2 \sum_{j=2}^3 \sum_{{\bf k},s} \hbar[
g_{j\,{\bf k},s} \, a_{{\bf k},s} \,  
{\rm e}^{i (\omega_{j1}-\omega_k) t} \, 
{\rm e}^{i {\bf k} \cdot {\bf r}_i} S_{ij}^+ \nonumber\\
 + \, {\rm H.c.} \, ] + H_L ~,
\end{eqnarray}
with the coupling constants 
\begin{eqnarray} \label{22}
g_{j\,{\bf k},s} &=& i e 
\left( {\omega_k \over 2 \epsilon_0 \hbar L^3} \right)^{1/2} 
\left( {\bf D}_{1j}, {\bf \epsilon}_{{\bf k},s} \right),
\end{eqnarray}
and laser part 
\begin{eqnarray} \label{23}
H_L &=& \frac{\hbar}{2} \, \sum^2_{i = 1} \sum^3_{j = 2} 
\Omega_{j} \left\{ S_{ij}^+ + S_{ij}^- \right\} .
\end{eqnarray}
The Rabi frequencies of the lasers are $\Omega_{j} = (e/  \hbar) \, 
{\bf D}_{1 j} \cdot {\bf E}_{0j}$ for $j=2,3$, and they are the same
for both atoms.  The operator $H_I$ implicitly 
contains the dipole-dipole interaction of the two atoms, 
as seen from the conditional Hamiltonian $H_{\rm cond}$ further 
below. In the  Power-Zienau formulation, which we have used above, 
this interaction is due to photon exchange \cite{aga}.

{\em Conditional Hamiltonian and waiting times}.
As explained in Refs. \cite{HeWi,Wi,He,HeSo,PleKni}, $H_{{\rm cond}}$ is
obtained (in the interaction picture) from the short-time development under 
the condition of no emission, i.e. from the relation
\begin{equation}
\label{24}
{\bf 1} - \frac{ i}{\hbar}~H_{{\rm cond}} \Delta t = \langle 0_{ph}|
U_I(\Delta t, 0) |0_{ph}\rangle
\end{equation}
where the right-hand side is evaluated in second order perturbation theory for
$\Delta t$ intermediate between inverse optical frequencies and atomic 
decay times. In a similar way as for a single atom \cite{HeWi,Wi,He,HeSo} 
one obtains for the system of two three-level atoms \cite{Be}
\begin{eqnarray} \label{25}
H_{\rm cond} &=& \frac{\hbar}{2i} \Big[ 
\sum_{j=2}^3 A_j \left( S_{1j}^+S_{1j}^- + S_{2j}^+S_{2j}^- \right)
\nonumber \\
&&~~~~~~~~~+ C_j \left( S_{1j}^+S_{2j}^- + S_{2j}^+S_{1j}^- \right) \Big] + H_L
\end{eqnarray}
with the $r$-dependent coupling constants 
\begin{eqnarray} \label{26}
C_j&=& {3A_j \over 2} \, {\rm e}^{i k_{j1} r} \Bigg[
{1 \over {i}k_{j1} r} \left( 1 - \cos^2 \vartheta_j \right) \nonumber\\
&& + \left( {1 \over (k_{j1}r)^2} -{1 \over i(k_{j1}r)^3} \right) 
\left( 1 - 3 \cos^2 \vartheta_j \right) \Bigg]~~~
\end{eqnarray}
which contain the dipole-dipole interaction between the atoms. The 
dependence of $C_j$ on $r$ is maximal for $\vartheta_j = \pi/2$ (see
Fig. 2). 
In the following we will assume for the Einstein coefficients and Rabi
frequencies the relations
\begin{eqnarray} \label{11}
\Omega_2 \ll \Omega_3 ~,~~ \Omega_2 \ll \Omega_3^2/A_3,
~~\mbox{and}~~ A_2 \approx 0~.
\end{eqnarray}
Then $A_2$ and  ${\rm Re} \, C_2 $ can be neglected in $H_{\rm
  cond}$; we will also neglect ${\rm Im} \, C_2$ which is allowed if
 $r$ is not small compared to $\lambda_{21}$, as seen from Fig. 2.

Let $|g \rangle$, $|e_2 \rangle$ and $|e_3 \rangle$ denote the states 
where both atoms are in the ground state and  the excited states $|2 \rangle$ 
and $|3 \rangle$, respectively, and let $|s_{jk} \rangle$  be the
symmetric and $i|a_{jk} 
\rangle$ be the antisymmetric combinations of $|j \rangle 
|k \rangle$ and $|k \rangle |j \rangle$. Then Eq.~(\ref{25}) becomes
\begin{eqnarray} \label{29}
H_{\rm cond} &=& {\hbar \over 2 i} \, \Big[ 
A_3 \, \big( |s_{23} \rangle \langle s_{23}|  
+ |a_{23} \rangle \langle a_{23}| \big)
+   (A_3 + C_3) |s_{13} \rangle \langle s_{13}| 
+ ( A_3 - C_3) |a_{13} \rangle \langle a_{13}| 
+ 2 A_3 \, |e_3 \rangle \langle e_3|\nonumber \\
&&+ \Big\{\sum_{j=2}^3 \sqrt{2} i \Omega_j \big( |g \rangle
  \langle s_{1j}|  
+ |s_{1j} \rangle \langle e_j| \big) 
+  i \Omega_2 \big( |s_{13} \rangle \langle s_{23}| 
+ |a_{13} \rangle \langle a_{23}| \big) 
+  i \Omega_3 \big( |s_{12} \rangle \langle s_{23}| 
- |a_{12} \rangle \langle a_{23}| \big) + {\rm H.c.}\Big\} \Big] .
\end{eqnarray}
Without  lasers the conditional Hamiltonian is diagonal in this 
basis. 

Between emissions the atomic time development is given by $ U_{{\rm cond}} 
(t, 0) = \exp \left\{ - i H_{{\rm cond}} t/\hbar \right\} $
which is non-unitary since $H_{{\rm cond}}$ is non-hermitian. The
corresponding decrease in the norm of a vector is connected to the
waiting time distribution \cite{CohDal} for emission of a (next) photon. 
If at $t = 0$ the initial atomic state is $| \psi \rangle$ then the 
probability $P_0$ to observe {\em no} photon by a broadband detector 
(over all space) is given by \cite{HeWi,Wi,He,HeSo}
\begin{equation} \label{27}
P_0 (t;  |\psi \rangle) = \|U_{{\rm cond}}(t, 0) |\psi \rangle \|^2~,
\end{equation}
and the probability density $w_1$ of finding the first photon at time
$t$ is 
\begin{equation} \label{28}
w_1 (t; |\psi \rangle) = -~\frac{d}{dt}~P_0 (t; |\psi \rangle)~.
\end{equation}
For an initial density matrix instead of $|\psi \rangle$ the expressions 
are analogous, with a trace instead of a norm squared in Eq. (\ref{27}).

 According to Eqs. (\ref{27}) and (\ref{28}), 
$A_3 \pm {\rm Re} \, C_3$ describe the decay rates of 
$|s_{13} \rangle$ and $|a_{13} \rangle$, respectively. From this the
well-known fact  
follows that two atoms with dipole interaction can decay faster or slower 
than two independent atoms (superradiance and subradiance \cite{Brewer1}). 
${\rm Im} \, C_3$ corresponds to a level shift of $|s_{13}\rangle$ and
$-{\rm Im} \, C_3$ to a level shift of $|a_{13}\rangle$, caused by  
the interaction between the atoms.

{\em Reset matrix}.
Now we determine the reset operation which gives the state or density 
matrix right after a photon detection. Let  the state of the 
combined system, atoms plus quantized radiation field, be given at
time $t$ by $ |0_{\rm ph} \rangle\, \rho\,\langle 0_{\rm ph}| $,
i.e. the atomic system is described by the density matrix $\rho$ and
there are no photons (recall that the laser field is treated
classically). If at time $t + \Delta t$ a photon is found by a
non-absorptive measurement the combined system is in the state
\begin{equation}\label{210}
I\!\!P_{\!>} U_I(t + \Delta t, t) |0_{\rm ph} \rangle \,\rho \,\langle
0_{\rm ph}|
U_I^\dagger (t + \Delta t, t) I\!\!P_{\!>}
\end{equation}
where $ I\!\!P_{\!>} = {\bf 1} - |0_{\rm ph} \rangle {\bf 1}_A 
\langle 0_{\rm ph}| $
is the projector onto the one or more photon space (since $\Delta t$
is in the above range and thus small 
one could directly take the projector onto the one-photon
space). The probability for this event is the trace over Eq.
(\ref{210}). For the state of the atomic system it is irrelevant
whether the detected photon is absorbed or not (intuitively the photon
travels away and no longer interacts with the atomic
system). Hence after a photon detection at time $t + \Delta t$ the
non-normalized state of the atomic system alone, denoted by ${\cal R} 
(\rho) \Delta t$, is given by the partial trace over the photon space, 
\begin{eqnarray}\label{211}
&&{\cal R} (\rho) \Delta t =  \nonumber \\
&& {\rm tr_{ph}} \left( I\!\!P_{\!>} U_I (t
+ \Delta t,t) |0_{ph} \rangle \,\rho \,\langle0_{ph}|U^\dagger_I (t +
\Delta t, t)
 I\!\!P_{\!>} \right).
\end{eqnarray}
We call ${\cal R} (\rho)$ the non-normalized reset state \cite{He}. 
Proceeding as in Refs. \cite{He,HeSo} and  using
perturbation theory one obtains \cite{Be}
\begin{eqnarray} \label{212}
{\cal R} (\rho) &=& {\rm Re} \, C_3 
\left( S_{13}^-\rho S_{23}^+ + S_{23}^-\rho S_{13}^+ \right) \nonumber
\\
&& + A_3 \left(S_{13}^- \rho S_{13}^+ +S_{23}^- \rho S_{23}^+ \right) . 
\end{eqnarray} 
The normalized reset state is $\hat{{\cal R}}(\rho) \equiv {\cal R}
(\rho)/{\rm tr} {\cal R}(\rho)$. By Eq. (\ref{210}) the normalization of 
${\cal R}(\rho)$ is such that tr$_A {\cal R}(\rho) \Delta t$ is the 
probability for a photon detection at time $t + \Delta t$ when the 
(normalized) state of the atomic system at time $t$ is $\rho$. 
The laser field does not appear in the reset state, just as in the
case of a single atom \cite{He,HeSo}, since its effect 
during the short time $\Delta t$  is negligible.

By a simple calculation one checks that Eq. (\ref{212}) can be written as 
\begin{equation} \label{213}
{\cal R} (\rho) = \left(A_3 + {\rm Re} \, C_3 \right) R_+ \rho R^\dagger_+ 
+ \left(A_3 - {\rm Re} \, C_3 \right) R_- \rho R_-^\dagger
\end{equation}
where
\begin{eqnarray} \label{214}
R_+ &=& \left(S_{13}^- + S_{23}^- \right) / \sqrt{2}  \nonumber\\ 
&=& |g \rangle \langle s_{13}| + |s_{13} \rangle \langle e_3|
+ \big(|s_{12} \rangle \langle s_{23}| - |a_{12} \rangle \langle a_{23}| 
\big)/\sqrt{2}  \nonumber \\
R_- &=& \left(S_{13}^- - S_{23}^- \right) / \sqrt{2} \nonumber \\
&=&    |g \rangle \langle a_{13}| + |a_{13} \rangle \langle e_3| 
+ \big(|s_{12} \rangle \langle a_{23}| + |a_{12} \rangle \langle
s_{23}| \big)/\sqrt{2} 
~.
\end{eqnarray}
If $\rho$ is a pure state, $\rho = | \psi \rangle \langle \psi |$
say, then $R_\pm \rho R_\pm^\dagger$ are also pure states. This
decomposition of ${\cal R}(\rho)$ is advantageous for simulations of
trajectories. As pointed out above, $A_3 \pm{\rm Re}\, C_3$ describe
the decay rates of 
$|s_{13} \rangle$ and $|a_{13} \rangle$ to $|g \rangle$. The state 
$|e_3 \rangle$ can decay  both to $|s_{13} \rangle$ and $|a_{13} \rangle$, 
with respective decay rates $A_3 \pm {\rm Re}\, C_3$. The decay rate 
of the states $|s_{23} \rangle$ and $|a_{23} \rangle$ is $A_3$ and is
the same as in the case of two independent atoms.

{\em Simulation of a single trajectory.}
Starting at $t = 0$ with a pure state, the state develops
according to $U_{{\rm cond}}$ until the first emission at some time
$t_1$, determined from $w_1$ in Eq. (\ref{28}). Then the state is
reset according to Eq. (\ref{212}) to a new density matrix (which has
to be normalized), and so on.

The decomposition of ${\cal R}(\rho)$ in Eq. (\ref{213}) allows one,
however, to work solely with pure states which is numerically much
more efficient. One can start with a pure  state $| \psi \rangle$,
develop it with $U_{{\rm cond}}$ until $t_1$ to the (non-normalized)
$|\psi(t_1)\rangle$, reset to one of the pure states
$ R_\pm |\psi(t_1)\rangle/\| \cdot \| $
with relative probabilities given by the factors $A_3 \pm {\rm Re} \, C_3$
appearing in Eq. (\ref{213}), and so on. The waiting time distributions 
are not changed by this procedure.
\section{ Fluorescence jumps for two   atoms}
For a single atom in a $V$ configuration the existence of dark
periods   is due to two widely
different time scales in the times between two
subsequent photon emissions (cf., e.g., Refs. \cite{CohDal,HeWi,Wi,HePle1}).
The smaller time scale, $T'$, is of the order of $A_3^{-1}$, while the
larger time scale, $T''$, is the inverse of the
smallest eigenvalue of $H_{\rm cond}/i$ for a single atom.
 One can pick a time,
$\tilde T$, say, with $T' \ll \tilde T \ll T''$, and if the time
between two subsequent photon emissions is longer than $\tilde T$ one
may then define this as a dark period. The mean duration of such dark
periods is essentially independent of $\tilde T$, if  chosen as above, and
is given by $T''$. When the waiting time between two photons is less
than $\tilde T$ the atom is said to be in a light period. The average
intensity in a not too short light period is that of a driven two-level
system, i.e. in our case the 1 and 3 levels.

If one has two independent, {\em noninteracting}, atoms  the 
combined fluorescence is just the sum of the individual
contributions. When both atoms are in a dark period one has a dark
period of the combined system. If only one atom is in a dark period
one observes a fluorescence period with intensity of that of a
single two-level atom (single-intensity period), and if both atoms are
radiating one observes a double-intensity period. However, due to
fluctuations in the emission times the latter two periods are not
sharply defined if the atoms are so close to each other that one
cannot determine from which atom a particular 
photon came. To distinguish the periods one has therefore to use an 
average photon intensity,  obtained by means of an 
averaging time $\Delta T$. This $\Delta T$
 has to be large enough so that the photon intensity doubles when both
 atoms are not  in a dark period. If, on the other hand, $\Delta T$ is
 chosen too large one may overlook short fluorescence periods, and one
 will see more seemingly direct 
transitions between dark periods and double intensity
 periods (double jumps). The analytic treatment of fluorescence of
 two independent atoms is easily obtained from the single-atom case
 \cite{Sauter}.

For two {\em dipole-interacting} atoms which are sufficiently close to
each other the photons cannot be attributed to a particular
atom, either,  and one has to consider the two interacting atoms as a jointly
radiating system. To be able to differentiate between different
fluorescence phases one again has  to average the photon numbers over a
time interval $\Delta T$ to arrive at an intensity, 
and in the following we present the results of numerical simulations which
have been obtained by the methods explained in Section II.

{\em Atomic distances of a few wavelengths}. 
Fig. 3 shows the number of emitted photons per time $A_3^{-1}$, 
averaged over a time
interval  $\Delta T = 190/A_3$. If the
atomic distance is larger than a third of a wavelength of the fast
transition, $r >\frac{1}{3} \lambda_{31}$, i.e. $k_{31} r > 2$, one can clearly
distinguish three types of fluorescence periods,  dark periods (0),
single-intensity periods (1), and double-intensity  periods (2). For two
noninteracting atoms these would correspond to radiation of no atom,
one atom, and two atoms, respectively. However, in the case of
interaction  and small distance the 
system of two atoms radiates as a whole, and in general one cannot
attribute periods 1 and 2 to radiation of individual atoms as in the
noninteracting case.

The transition between the periods occurs rapidly but not
instantaneously. The duration of the periods is long compared to the
atomic time scale. From a sufficiently long trajectory one can obtain
the average lengths of the periods, denoted by $T_0, T_1$ and
$T_2$. The results are shown in Fig. 4. The Rabi frequency $\Omega_2$
of laser 2, which drives the weak atomic 1 - 2 transitions, has been chosen
in such a way that for independent atoms one has $T_0 = 2000/A_3$. As
seen in Fig. 4, $T_0$ is essentially independent of the atomic
distance and thus of the dipole-dipole interaction. In contrast to
this the two light periods are strongly distance dependent. In Fig. 4
(a), $T_2$ varies between $1200/A_3$ and $2600/A_3$ and a similar
behavior is also seen in Fig. 4 (b). The curve for $T_1$ resembles
that for $T_2$, except for smaller relative variation.

There is an interesting correspondence between the $r$ dependence of
$T_1$ and $T_2$ with that of ${\rm Re} \, C_3$. As seen from a comparison of
Figs. 4 and 5 the variations with $r$ of all three quantities seem to
be in phase. For ${\rm Im} \, C_3$ the variation is out of phase, as
seen from Eq. (\ref{26}). Since ${\rm Re} \, C_3$ influences the decay
rates of the two-atom 
systems, this in-phase behavior suggests that the variation in the
lengths of periods 1 and 2 are due to an $r$ dependence of the decay
rates. 

{\em Small atomic distances}. Our simulations have shown furthermore
that for $r <\frac{1}{4} \lambda_{3 1}$ the intensity in
period 2 decreases and no longer reaches that of two simultaneously
radiating independent atoms. For very small distance only periods 0
and 1 remain. The reason for this will be discussed below.
\section{Interpretation of results}
For a single atom (in a $V$ configuration as in Fig. 1) one can associate
light and dark periods with certain atomic states and density
matrices. During a dark period the atomic state rapidly approaches the
eigenstate of $H_{{\rm cond}}$ with smallest imaginary part of the
eigenvalue \cite{HeWi,HePle1,BeHe1}. This eigenstate is very close to $| 2
\rangle$, up to terms of order $\Omega_2 A_3/\Omega_3^2$ and
$\Omega_2/\Omega_3$ \cite{BeHe1}. Thus, in a dark period, the atom can
be regarded to be approximately in the state $| 2 \rangle$. During a
light period the atom can be regarded to be in the equilibrium state
(density matrix) of the 1 - 3 subsystem driven by $\Omega_3$, again up
to terms of the above orders \cite{BeHe1}. A jump from one
fluorescence period to the other corresponds to a transition between
these atomic states, and such a transition is caused by laser 2. With
$\Omega_2 = 0$ and $A_2 = 0$ there would be no
transitions. This correspondence clearly carries over to the three
fluorescence periods of two independent atoms.

We are now going to suggest a similar correspondence between
fluorescence periods and states for two dipole-interacting atoms. Fig. 6
depicts the Dicke states of the two-atom system (see
Eq. (\ref{29})). Dashed arrows indicate the weak driving by laser 2,
solid arrows indicate strong driving by laser 3 and decay,
respectively. Now, for $\Omega_2 = 0$, i.e. no dashed arrows, the
states in Fig. 6 decompose into three non-connected
subspaces, namely one spanned by $|e_2 \rangle$ and the two others
spanned by the four inner  and  outer states, respectively:
\begin{eqnarray*}
{\rm subspace}~~ 0 & : & |e_2 \rangle\\
{\rm subspace}~~ 1 & : & |s_{1 2} \rangle, |a_{1 2} \rangle, |s_{2 3}
\rangle, |a_{2 3}\rangle\\
{\rm subspace}~~ 2 & : & |g \rangle, |s_{1 3} \rangle, |a_{1 3}
\rangle, |e_3 \rangle
\end{eqnarray*}
If the two atoms are in state $|e_2 \rangle$ then each of them is in
its dark state, and thus no photon can be emitted. For $\Omega_2 = 0$,
the time development in subspace 1 is  exactly the same as
that for a system of two noninteracting atoms in the same subspace of
states. This can be seen directly from Eq. (\ref{29}) 
 and it is also physically obvious since
two atoms can only interact via photon exchange if none of them is in
the dark state $|2 \rangle$. The photon rate for subspace 1 is
therefore that of a single  two-level atom with levels 1 and
3. Subspace 2 corresponds to the level scheme of a system of two
dipole-dipole interacting two-level atoms (with levels 1 and 3), as
for example recently discussed in Ref. \cite{BeHe4}. The photon rate
of this system is, in good approximation, twice that of a single
two-level atom, provided $r > ~\frac{1}{4}~\lambda_{3 1}$. For smaller
atomic distance the photon rate rapidly decreases to zero, due to the
increasing level shifts.

If the two-atom system has been in one of the subspaces 0, 1, or 2, it
will quickly approach the corresponding equilibrium state for driving
$\Omega_3$ and distance $r$.  If
$\Omega_2 \not= 0$ the additional weak driving will, from time to
time and in analogy to a single V system, cause transitions between
the three subspaces 0, 1, and 2, and 
each transition will correspond to a jump in the fluorescence.

Thus the fluorescence periods 0, 1, and 2 should correspond to (the
equilibrium states of) the subspaces 0, 1, and 2, respectively. This
is verified by the numerical evaluation in Fig. 7. In the
lower part of Fig. 7 a particular realization of an
intensity trajectory with alternating periods of fluorescence is
plotted. The three upper curves show the populations of the three
subspaces corresponding to this realization, obtained by the
conditional Hamiltonian and reset matrix of Section II. The agreement
between fluorescence periods 0, 1, 2, and subspaces 0, 1, 2 is
striking. During dark periods the two-atom system is in the subspace
0, and similarly for periods 1 and 2.

This correspondence, however, depends to some extent on how large
$\Delta T$ is chosen for the averaging of photon counts. If $\Delta T$
is chosen too large one can overlook some jumps between subspaces and some
very short fluorescence periods. If 
$\Delta T$ is chosen too small there may be large intensity
fluctuations, resulting in an incorrect determination of the different
periods.

If the atomic distance decreases below $\frac{1}{4}\lambda_{31}$
the level shifts of $|s_{13}\rangle$ and $|a_{13}\rangle$ by 
${\rm Im} \, C_3$ increase rapidly. This renders the driving by laser
3 within the subspace 2 much less efficient. Hence for very small
atomic distances the driving is essentially restricted to subspace
1. This explains the vanishing of double-intensity periods for very small
atomic distances.
\section{Discussion and summary}
Cooperative effects of driven two three-level atoms have been studied,
where each individual atom can exhibit light and dark periods. 
The atoms were considered to be a fixed distance $r$
apart. If $r$ is of the order of a few wavelengths of the fluorescent
light the individual
photons are no longer attributable to a particular atom and the
two-atom system radiates as a whole, due to the 
dipole-dipole interaction. In addition to dark periods the system
shows two types of light periods, one with fluorescence intensity as
if only a single atom were 
radiating, and the other with double intensity.

We have proposed to study the mean duration, $T_1$ and $T_2$, of the
two types of light periods as a quantity sensitively depending on the
dipole-dipole interaction and thus on the atomic
distance. Experimentally and numerically these quantities are easily
accessible.

 We have performed fluorescence simulations for atomic separations of up to 5
wavelengths and have found oscillations in $T_1$ and $T_2$ of up to
40\% in amplitude. The amplitude decreases with the atomic separation
but  the oscillations seem to continue for separations larger than 5
wavelengths. 
By simulations we have shown that the $r$ dependence of $T_1$ and
$T_2$ is in phase with that of the real part of the dipole-dipole
constant. This is eminently reasonable since ${\rm Re} \, C_3$ directly
influence the decay rates of the excited states of the two-atom
system.

We have associated the three types of fluorescence periods with certain
subspaces of states for the two atoms and to equilibrium states in
these subspaces. The equilibrium states depend on the driving of the
strong transition and on the distance. The weak driving then causes
transitions between the subspaces. The transition rates depend not
only on the weak driving, but also on the form of the respective
equilibrium states and thus on the strong driving and on the atomic
distance. In contrast to the mean durations of the light periods the
mean duration of the dark periods is practically independent of
$r$. This is intuitively quite clear since in the dark state there is
essentially no photon exchange and thus no induced dipole-dipole interaction.

To define fluorescence periods one has to average the number of photon
emissions over a time interval $\Delta T$ of a some finite length.
Hence very short fluorescence periods are washed out and not observed,
and this can lead to
apparent direct transitions between double-intensity periods and dark
periods, or vice versa, so-called double jumps. Experimentally these have
been seen in Ref. \cite{SauterL}.
The cooperative effects of up to 40\% found by us
in the duration of the single-intensity and the double intensity periods
are noticeable.  Therefore we expect that also the frequency
for the appearance of double jumps is modified by the dipole-dipole  
interaction, but we cannot predict whether the changes are of the two  
orders of magnitude reported in Ref. \cite{SauterL}. Besides, the
system in Ref. 
\cite{SauterL} differs from the one considered here, and it is not
obvious if and 
how our results would carry over to that system.

\newpage
\centerline{FIGURE CAPTIONS} 

\noindent
FIG. 1. V system with metastable level 2 and Einstein
coefficient $A_3$ for level 3. $\Omega_2$ and $\Omega_3$ are the Rabi
frequencies of the two lasers driving the weak 1-2 transition and the
strong 1-3 transition, respectively.
\\[1cm]
FIG. 2. Dependence of $C_j/A_j$ $(j=2,3)$ on $r$
\\[1cm] 
FIG. 3. Number of photons, $I(t)$,  per time $A_3^{-1}$ emitted by 
two dipole-interacting atoms, averaged over  $\Delta T=190/A_3$, for 
 $\Omega_2=0,01\,A_3$, 
  $\Omega_3=0,5\,A_3$,  and $\vartheta_3=\pi/2$.
  (a) $k_{31}r=10$, (b) $k_{31}r=5$, and (c) $k_{31}r=2$. 
  The dashed curve indicates the fluorescence type (zero, single, and double
  intensity).
\\[1cm]
FIG. 4. $T_0$, $T_1$ and $T_2$ as a function of $r$ for
  $\vartheta_3=\pi/2$ and $\Delta T=250/A_3$.
  (a) $\Omega_3=0.3\,A_3$ and (b) $\Omega_3=0.6\,A_3$. $\Omega_2$ has
  been chosen such that for two noninteracting atoms one has
 $T_0 = 2000/A_3$.
\\[1cm]
FIG. 5. ${\rm Re} \, C_3/A_3$ as a function of $r$ for 
  $\vartheta_3=\pi/2$.
\\[1cm]
FIG. 6.  Dicke states. The dashed and solid double arrows denote
  weak and strong driving, respectively. Simple arrows denote decays.
\\[1cm]
FIG. 7  Correspondence between fluorescence types and subspaces 
  0, 1, and 2 for  $\Omega_2=0,01\,A_3$, $\Omega_3=0,5\,A_3$ and 
  $\vartheta_3=\pi/2$. The atomic distance is chosen as $5
  \lambda_{31}/2\pi$.

\newpage
%\begin{center} FIG. 1 \end{center}
%\begin{minipage}{6.54truein}
\begin{figure}[htb]
\begin{center}
%\epsfxsize20.0cm
\centerline{\epsfbox{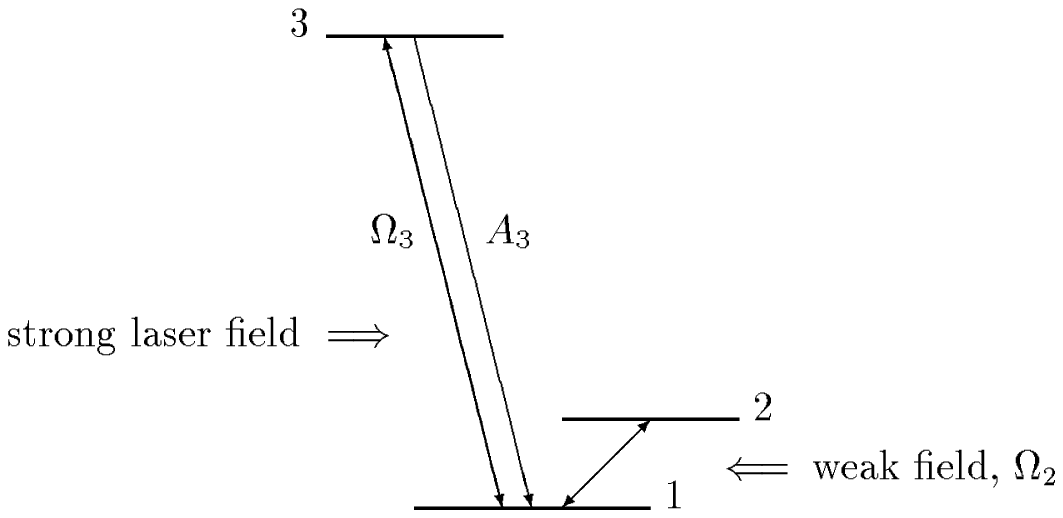}}
\centerline{\epsfbox{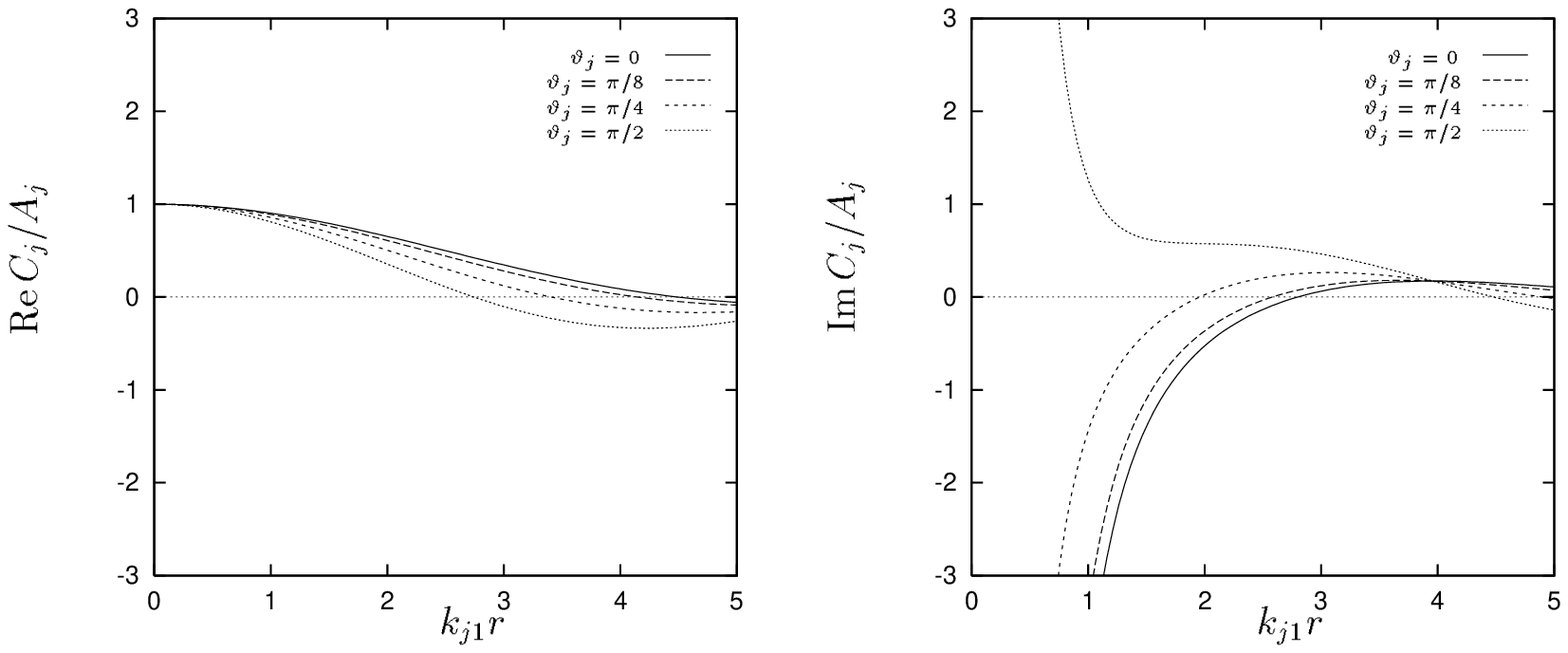}}
\centerline{\epsfbox{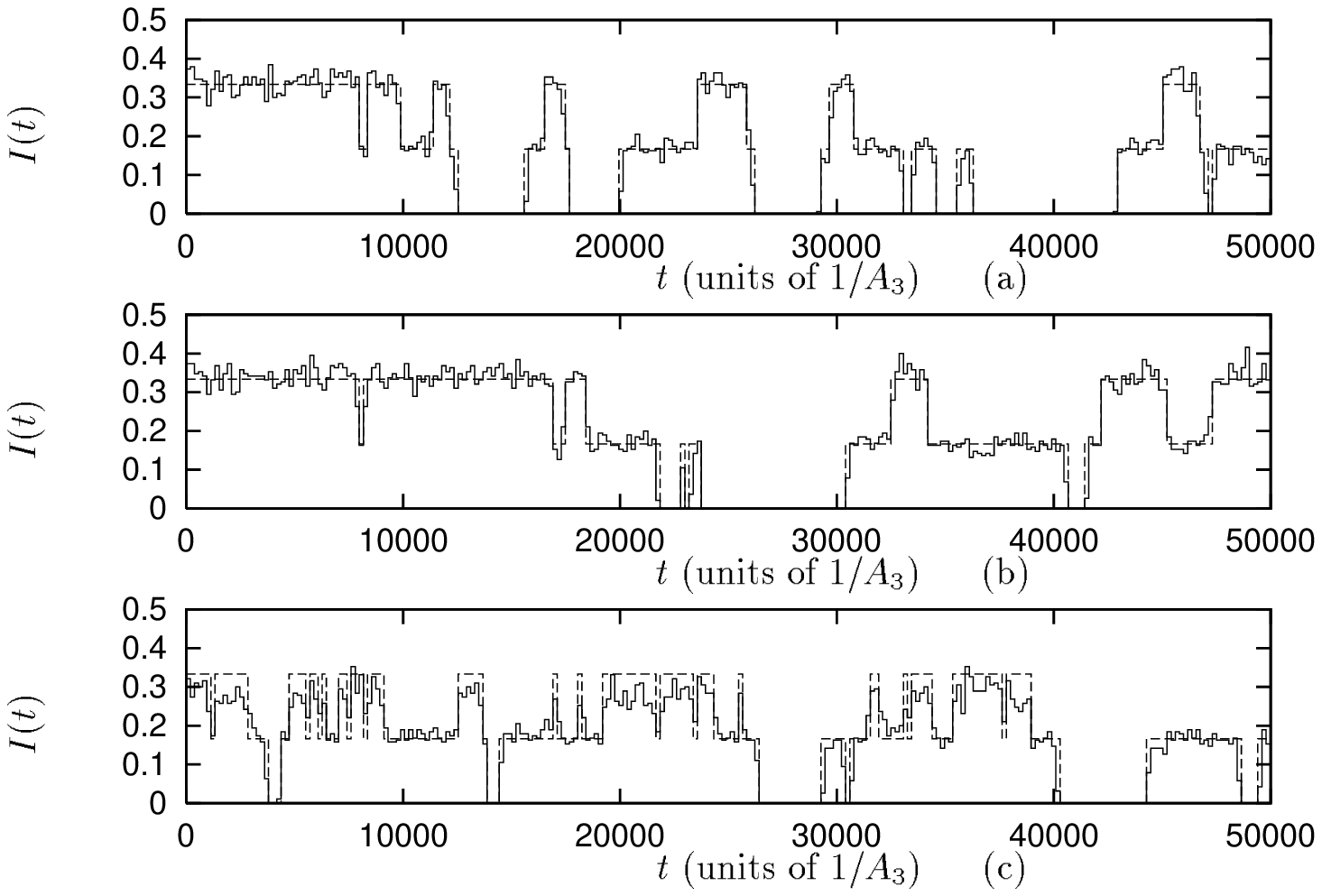}}
\centerline{\epsfbox{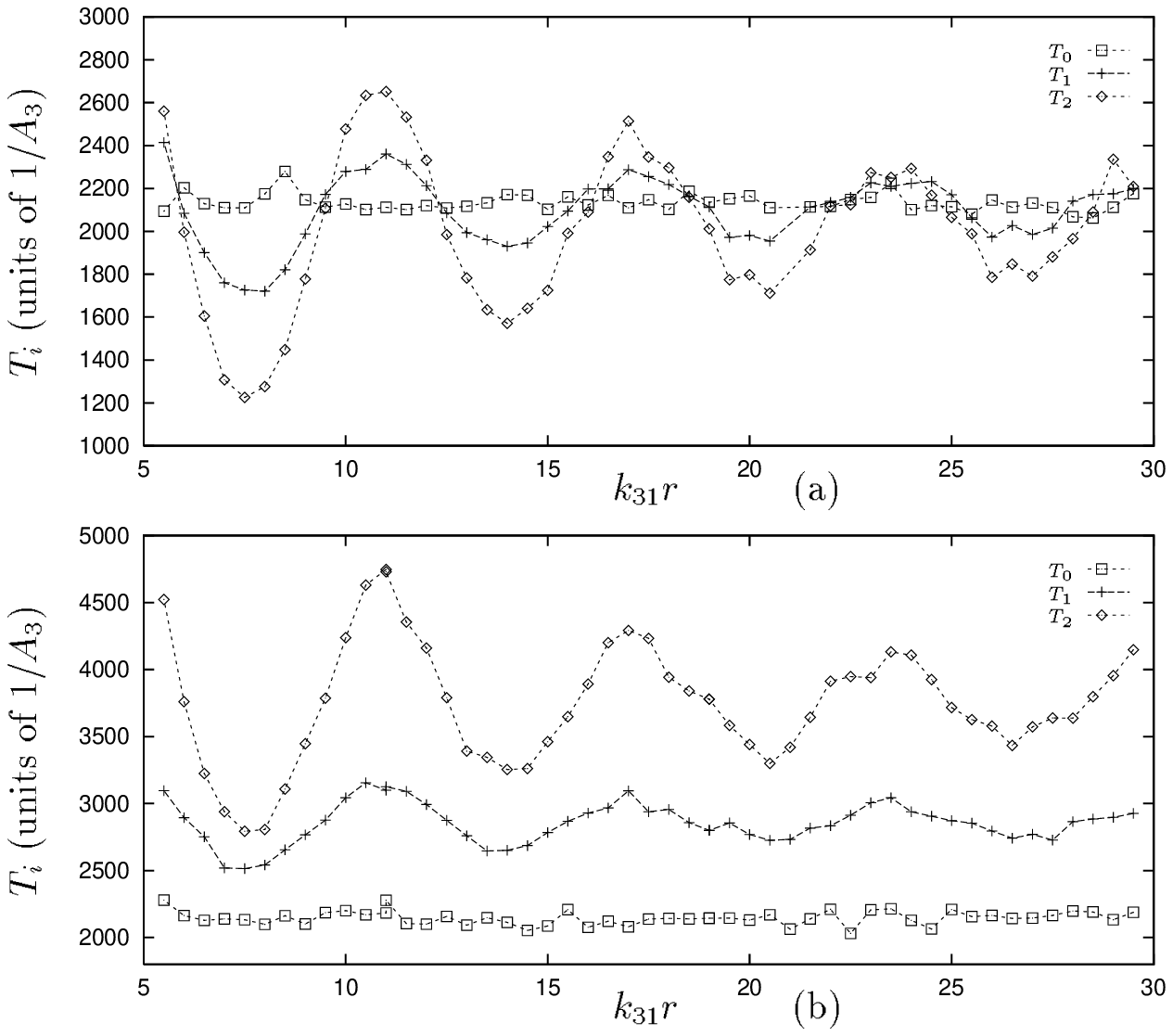}}
\centerline{\epsfbox{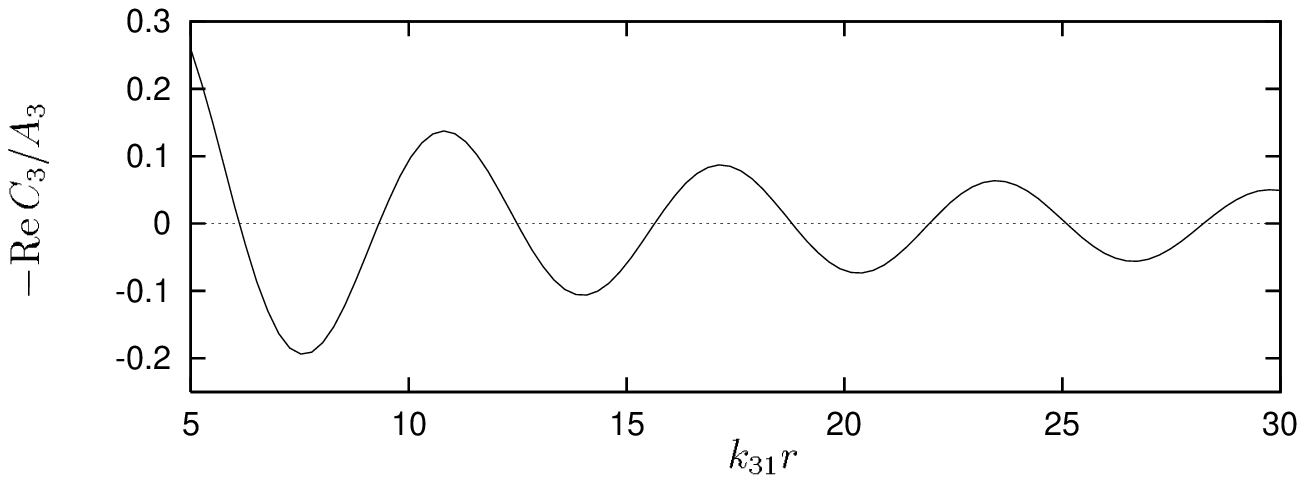}}
\centerline{\epsfbox{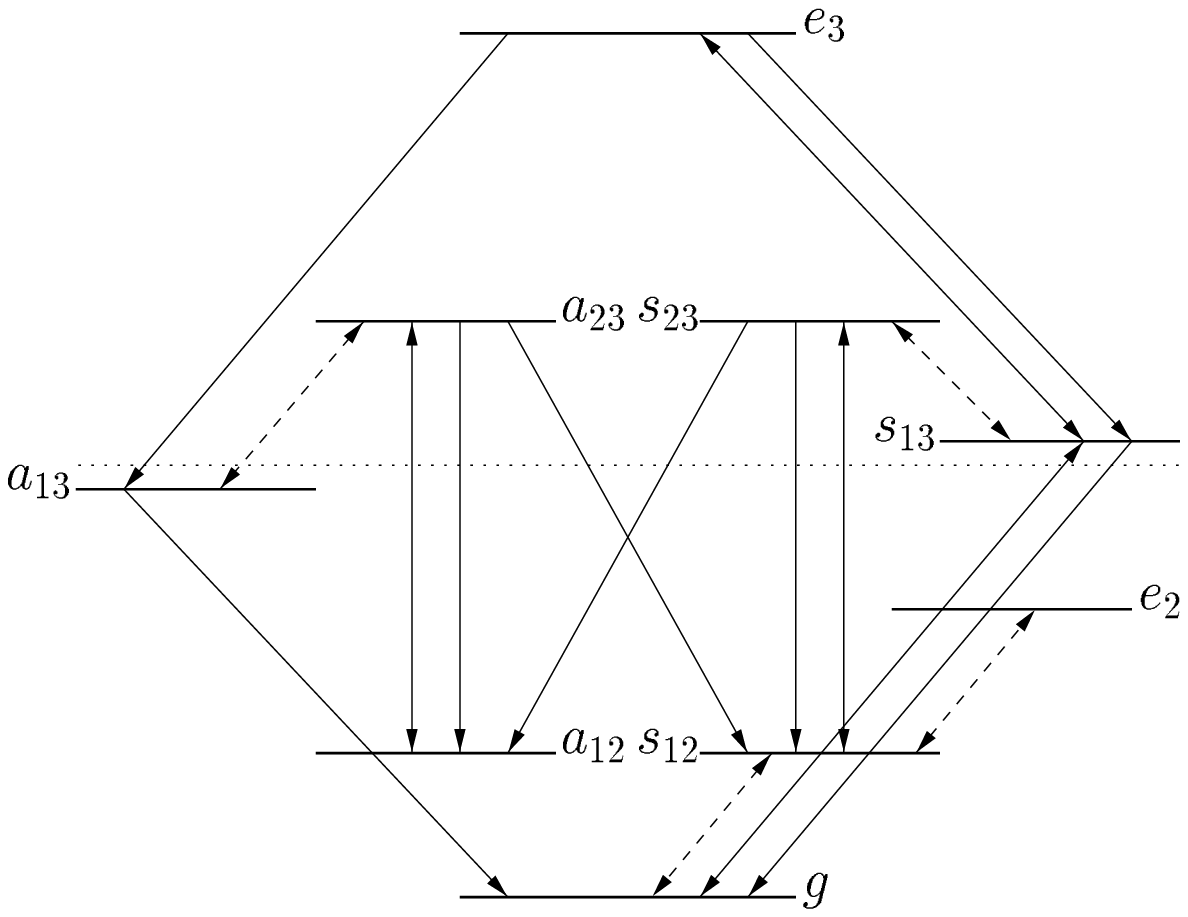}}
\centerline{\epsfbox{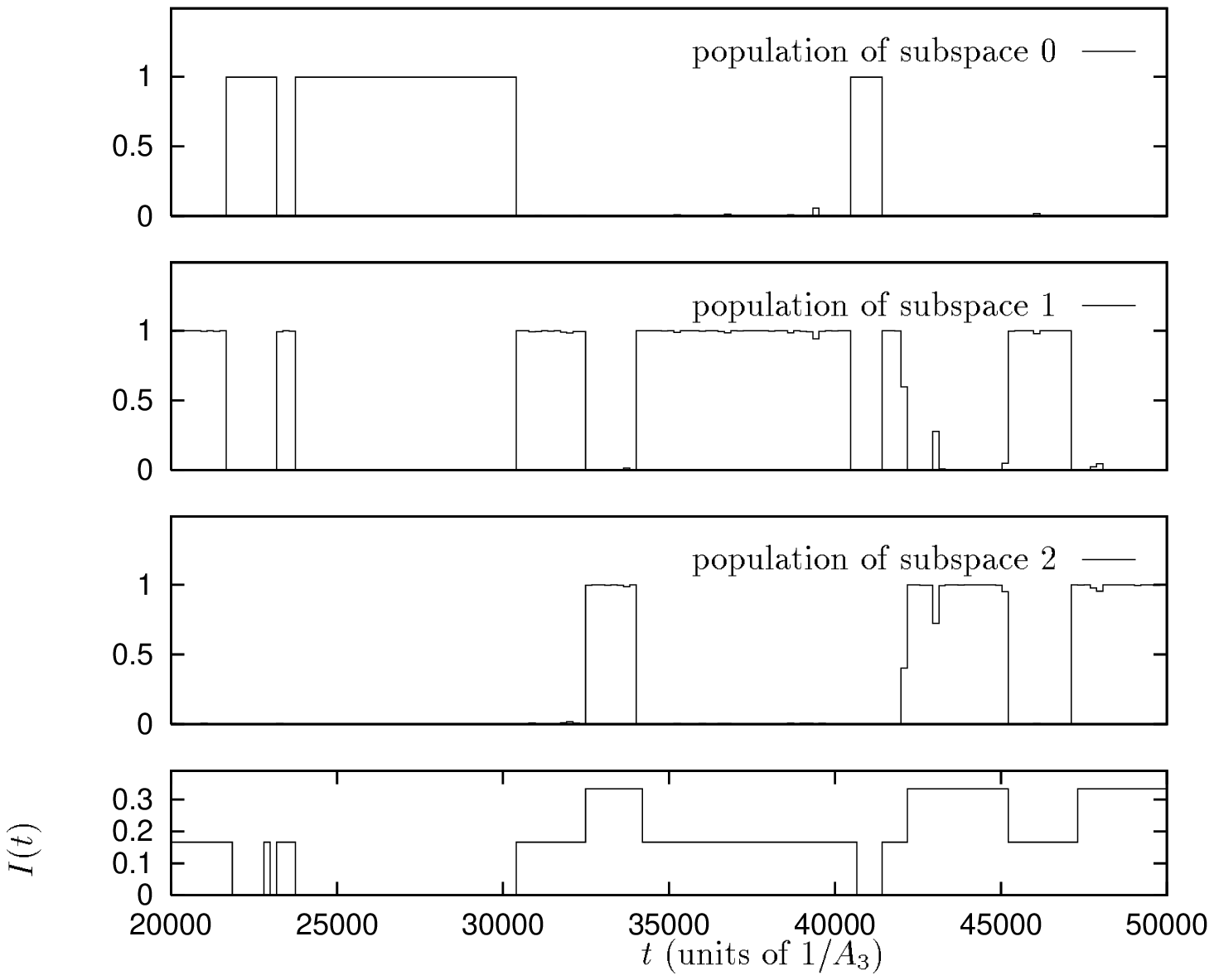}}
\end{center}
%\caption{} 
\end{figure}
%\end{minipage}


\begin{references}
\bibitem{rem1} In the experiments somewhat   different level schemes
  were considered.
\end{references}
\end{document}